\documentclass[10,conference]{IEEEtran}
\usepackage{graphicx}
\usepackage{cite}
\usepackage{float}
\usepackage{amsmath}
\usepackage{amsthm}
\usepackage{mathtools}
\usepackage{amssymb}
\usepackage{commath}
\usepackage{subfigure}
\usepackage{bbm}
\usepackage[justification=centering]{caption}
\usepackage[margin=1.9cm,nohead]{geometry}
\usepackage{color}

\newtheorem{definition}{Definition}

\IEEEoverridecommandlockouts

\ifCLASSINFOpdf
\else
\fi

\begin{document}
\title{The Stability Region of the Two-User Broadcast Channel}
\author{\IEEEauthorblockN{Nikolaos Pappas\textsuperscript{*}, Marios Kountouris\textsuperscript{\dag},\\}
\IEEEauthorblockA{\textsuperscript{*}Department of Science and Technology,
 Link{\"o}ping University, \\ Campus Norrk{\"o}ping, 60 174, Sweden\\
}
\IEEEauthorblockA{\textsuperscript{\dag}Mathematical and Algorithmic Sciences Lab, France Research Center, Huawei Technologies Co. Ltd.\\
Boulogne-Billancourt, 92100, France\\}
Emails: nikolaos.pappas@liu.se, marios.kountouris@huawei.com
\thanks{This work has been supported by the People Programme (Marie Curie Actions) of the European Union's Seventh Framework Programme FP7/2007-2013/ under REA grant agreement no.[612361] -- SOrBet.}}

\maketitle

\begin{abstract}
In this paper, we characterize the stability region of the two-user broadcast channel. First, we obtain the stability region in the general case. Second, we consider the particular case where each receiver treats the interfering signal as noise, as well as the case in which the packets are transmitted using superposition coding and successive decoding is employed at the strong receiver.
\end{abstract}

\section{Introduction}
A fundamental question to which Information Theory aims to provide an answer is how to maximize the use of a communication channel between transmitters and receivers. In other words, its major objective is to characterize the maximum achievable rate of information that can be reliably transmitted over a communication channel, which is called the channel capacity. In contrast to point-to-point channels, if the channel is shared among multiple nodes (multiuser channel), the goal is to find the capacity region, i.e. the set of all simultaneously achievable rates. One of the main assumptions in the information-theoretic formulation of the capacity region is that the maximum achievable rate is obtained under infinitely backlogged users. However, the bursty nature of the sources in communication networks gave rise to the development of a different concept of ``capacity region", which is the maximum stable throughput region or the \emph{stability region} \cite{rao:stability}.
Understanding the relationship between the information-theoretic capacity region and the stability region has received considerable attention in recent years and some progress has been made primarily for multiple access channels. Interestingly, the aforementioned regions (capacity and stability) are not in general identical and general conditions under which they coincide are known only in very few cases \cite{b:EphremidesHajekUnion}.

In this work, we consider the two-user broadcast channel \cite{b:CoverBroadcast75}, which models the simultaneous communication of information (different messages) from one source to multiple destinations. Marton in \cite{b:Marton} derived an inner bound, which is the best known achievable information-theoretic capacity region for a general discrete memoryless broadcast channel. Fayolle et al. \cite{b:FayolleBroadcast} provided a theoretical treatment of some basic problems related to the packet switching broadcast channel. The work in \cite{b:Jafarian} provided a partial characterization of the capacity region of the two-user Gaussian fading broadcast channel. 
Caire and Shamai in \cite{b:CaireShamaiToIT2003} investigated the achievable throughput of a multi-antenna Gaussian broadcast channel. In \cite{b:ZhouWunderAllerton2008}, scheduling policies in a broadcast system were considered and general conditions covering a class of throughput optimal scheduling policies were obtained. In \cite{b:JafarBC_stability}, the authors characterized the stability regions of two-user Gaussian fading multiple access and broadcast networks with centralized scheduling under the assumption of infinite backlogged users. In \cite{b:EphremidesBroadcast13}, the capacity region of the two-user broadcast erasure channel was characterized and algorithms based on linear network coding and their stability region were also provided.

Superposition Coding (SC) \cite{b:CoverBroadcast75} is one of the fundamental building blocks in network information theory. The objective of SC is to simultaneously transmit two messages by encoding them into a single signal in two layers. The receiver with the ``better" (less noisy) channel, also named stronger receiver, can recover the signal on both layers by applying successive interference cancelation, while the other (weaker or ``worse") can decode the message on the coarse layer treating the message on the fine layer (interference) as noise. In \cite{b:HaenggiISIT10}, SC with conventional frequency division in a Poisson field of interferers was analyzed. 
Furthermore, in \cite{b:HaenggiTWC2012}, the authors provided a software-radio based design and implementation of SC. Their results show that SC can provide substantial spectral efficiency gains compared to orthogonal schemes, such as time division multiplexing. The stability region of the two-user interference channel was derived in \cite{b:pappasitw2013-ic}, where the case of successive interference cancelation was also considered. 
 
In this paper, the stability region of the two-user broadcast channel is obtained. We first provide the stability region for the general case as a function of success probabilities and afterwards we specialize our study considering two particular cases. The first case is when both receivers treat interference as noise. The second case is when superposition coding is employed and the user experiencing better channel uses a successive decoding scheme. Two simple transmit power allocation schemes are considered in the latter case: i) the assigned power remains fixed, and ii) the transmit power is adapted to the state of the queues. 
 
\section{System Model} \label{sec:model}
\begin{figure}[]
\centering
\includegraphics[scale=1.3]{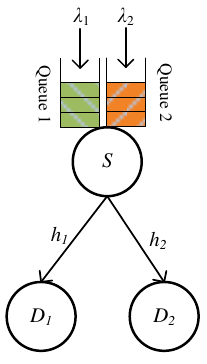}
\caption{The two-user broadcast channel with bursty arrivals.}
\centering
\label{fig:system_model}
\end{figure}

We consider a two-user broadcast channel, as depicted in Fig.\ref{fig:system_model}, in which a single transmitter having two different queues intends to communicate with two receivers. The first (resp. second) queue contains the packets (messages) that are destined to receiver $D_1$ (resp. $D_2$). Time is assumed to be slotted, the packet arrival processes at the first and the second queue are assumed to be independent and stationary with mean rates $\lambda_1$ and $\lambda_2$ in packets per slot, respectively. Both queues have infinite capacity to store incoming packets and $Q_i$ denotes the size in number of packets of the $i$-th queue. The source transmits packets in a timeslot if at least one of its queue is not empty. The transmission of one packet requires one timeslot and we assume that receive acknowledgements (ACKs) are instantaneous and error-free.

If only the $i$-th queue at the source is non-empty during a certain timeslot, then the transmitter sends information to the $i$-th receiver only. When both queues have packets, the source transmits a packet that contains the messages of both receivers, whereas whenever both queues at the source are empty, the transmitter remains silent.

Let $\mathcal{D}_{i/\mathcal{T}}$ denote the event that $D_i$ is able to decode the packet transmitted from the $i$-th queue of the transmitter given a set of non-empty queues denoted by $\mathcal{T}$, e.g. $\mathcal{D}_{1/1,2}$ denotes the event that the first receiver can decode the packet from the first queue when both queues are not empty ($\mathcal{T} = \{1,2\}$). It is evident that $\mathrm{Pr}\left(\mathcal{D}_{1/1} \right) \geq \mathrm{Pr}\left(\mathcal{D}_{1/1,2} \right)$.

The average service rate seen by the first queue is

{\scriptsize
\begin{eqnarray} \label{eq:mu1}
\mu_1 = \mathrm{Pr}\left(Q_2 > 0 \right) \mathrm{Pr}\left(\mathcal{D}_{1/1,2} \right) + \mathrm{Pr}\left(Q_2 = 0 \right) \mathrm{Pr}\left(\mathcal{D}_{1/1} \right).
\end{eqnarray}
}

Respectively, the average service rate of the second queue is

{\scriptsize
\begin{eqnarray} \label{eq:mu2}
\mu_2 = \mathrm{Pr}\left(Q_1 > 0 \right) \mathrm{Pr}\left(\mathcal{D}_{2/1,2} \right) + \mathrm{Pr}\left(Q_1 = 0 \right) \mathrm{Pr}\left(\mathcal{D}_{2/2} \right).
\end{eqnarray}
}

If a packet from the $i$-th queue fails to reach $D_i$, it remains in queue $i$ and is retransmitted in the next timeslot. 

The signal $y_i$ received at user $D_i$ at a timeslot $t$ is given by 
\begin{eqnarray} \label{eq:signal}
y_i^t = h_i^t x_i^t + n_i^t
\end{eqnarray}
where $n_i^t$ is the additive white Gaussian noise at timeslot $t$ with zero mean and unit variance. The channel gain from the transmitter to $D_i$ at time instant $t$ is denoted by $h_i^t$, and the transmitted signal is $x_i^t$. A block fading channel model with Rayleigh fading is considered here, i.e. the fading coefficients $h_i^t$ remain constant during one timeslot, but change independently from one timeslot to another. In the transmission phase (downlink), the transmitter assigns power $P_i$ for messages (packets) from queue $i$. 

The event $\mathcal{D}_{i/i}$ is defined as the probability that the uncoded received signal-to-noise ratio (SNR) is above a certain threshold $\gamma_i$, i.e. $\mathcal{D}_{i/i} = \left\lbrace \gamma_i \leq \mathrm{SNR}_i \right\rbrace$. The distance between the transmitter and $D_i$ is denoted by $d_i$ and $\alpha$ is the pathloss exponent. The SNR threshold for receiver $i$ is $\gamma_i$. Then $\mathrm{SNR}_i \triangleq |h_{i}|^2 d^{-\alpha}_{i} P_i$ assuming a physical layer model. The probability that the link between the transmitter and $D_i$ is not in outage when only the $i$-th queue is non-empty is given by (Ch. 5.4 in~\cite{b:Tse})
\begin{eqnarray} \label{eq:SNR}
\mathrm{Pr}\left(\mathcal{D}_{i/i}\right) =\mathrm{Pr} \left\lbrace \mathrm{SNR}_i \geq \gamma_i \right\rbrace = \exp \left(- \frac{\gamma_i d^{\alpha}_{i}}{P_{i}}\right).
\end{eqnarray}
Note that this is an approximation on the success probability under specific assumptions on the underlying physical layer model and is done in order to relate the success probabilities with a physical layer and channel model. Actually, the above expression on the success probability comes from the rates for arbitrarily reliable communication, which implies that the channel uses go to infinity. This (asymptotic) expression is an approximation of the instantaneous rate when information is transmitted in packets. Nevertheless, the main result of this paper, i.e. the stability region derived in the following section, is general as it is expressed in terms of success probabilities, which can in turn take on different expressions depending on the adopted physical channel or the information-theoretic model. 

When both queues at the source are non-empty at timeslot $t$ then the source transmits the signal $x^t=x_1^t +x_2^t$ where $x_i^t$ is the signal for the user $D_i$. Then, the signal $y_i$ received at user $D_i$ at a timeslot $t$ is given by $y_i^t = h_i^t x^t + n_i^t, i=1, 2,$ where $n_i^t$ is the additive white Gaussian noise at timeslot $t$ with zero mean and unit variance. The channel gain from the transmitter to $D_i$ is denoted by $h_i^t$ at instant $t$ and a Rayleigh block fading model is considered. In the transmission phase (downlink), the transmitter assigns power $P_i$ for messages (packets) of queue $i$ with $P_1+P_2=P$. We assume that each receiver $D_i$ knows each channel $h_i^t$ (perfect CSIR) and that the transmitter has perfect channel state information (CSIT), i.e. it knows $h_i^t$, $\forall i$. Each receiver $i$ decodes separately its message using the received signal $y_i$. 
The success probabilities in the case that both queues are non-empty depend on the interference handling technique and for that we study certain different cases in the following sections.

\subsection{Stability Criteria}

We use the following definition of queue stability~\cite{Szpankowski:stability}:

\begin{definition}
Denote by $Q_i^t$ the length of queue $i$ at the beginning of time slot $t$. The queue is said to be \emph{stable} if
$\lim_{t \rightarrow \infty} {Pr}[Q_i^t < {x}] = F(x)$ and $\lim_{ {x} \rightarrow \infty} F(x) = 1$. If $\lim_{x \rightarrow \infty}  \lim_{t \rightarrow \infty} \inf {Pr}[Q_i^t < {x}] = 1,$ the queue is \emph{substable}. If a queue is stable, then it is also substable. 
If a queue is not substable, then we say it is unstable.
\end{definition}

Loynes' theorem~\cite{b:Loynes} states that if the arrival and service processes of a queue are strictly jointly stationary and the average arrival rate is less than the average service rate, then the queue is stable. If the average arrival rate is greater than the average service rate, then the queue is unstable and the value of $Q_i^t$ approaches infinity almost surely. The stability region of the system is defined as the set of arrival rate vectors $\boldsymbol{\lambda}=(\lambda_1, \lambda_2)$ for which the queues in the system are stable.

\section{The Stability Region -- The General Case} \label{sec:stability_general}
In this section, we provide the stability region as a function of the success probabilities in the general case without considering specific interference handling techniques.

The average service rates of the first and second queue are given by (\ref{eq:mu1}) and (\ref{eq:mu2}), respectively.
Since the average service rate of each queue depends on the queue size of the other queue, it cannot be computed directly. Therefore, we apply the stochastic dominance technique~\cite{rao:stability}, i.e. we construct hypothetical dominant systems, in which one of the sources transmits dummy packets when its packet queue is empty, while the other transmits according to its traffic.

\subsection{First Dominant System: the first queue transmits dummy packets}
In the first dominant system, when the first queue empties, then the source transmits a dummy packet for the $D_1$, while the second queue behaves in the same way as in the original system. All other assumptions remain unaltered in the dominant system. Thus, in this dominant system, the first queue never empties, thus the service rate for the second queue is given by $\mu_2 = \mathrm{Pr}\left(\mathcal{D}_{2/1,2} \right)$.

Then, we can obtain stability conditions for the second queue by applying Loyne's criterion~\cite{b:Loynes}. The queue at the second source is stable if and only if $\lambda_2 < \mu_2$, thus $\lambda_2 < \mathrm{Pr}\left(\mathcal{D}_{2/1,2}\right)$.
Then we can obtain the probability that the second queue is empty by applying Little's theorem and is given by
\begin{eqnarray} \label{eq:Pr2empty_D1}
\mathrm{Pr}\left(Q_2 = 0 \right)  = 1-\frac{\lambda_2}{\mathrm{Pr}\left(\mathcal{D}_{2/1,2}\right)}.
\end{eqnarray}

After replacing (\ref{eq:Pr2empty_D1}) into (\ref{eq:mu1}), we obtain that the service rate for the first queue in the first dominant system is
\begin{eqnarray} \label{eq:mu1_D1}
\mu_1 = \mathrm{Pr}\left(\mathcal{D}_{1/1}\right) - \frac{\mathrm{Pr}\left(\mathcal{D}_{1/1}\right) - \mathrm{Pr}\left(\mathcal{D}_{1/1,2}\right)}{\mathrm{Pr}\left(\mathcal{D}_{2/1,2}\right)}\lambda_2.
\end{eqnarray}

The first queue is stable if and only if $\lambda_1 < \mu_1$. The stability region $\mathcal{R}_1$ obtained from the first dominant system
is given in (\ref{eq:R_1}).

\begin{figure*}
\begin{align} \label{eq:R_1}
\mathcal{R}_1 = \left\lbrace (\lambda_{1},\lambda_{2}): \frac{\lambda_1}{\mathrm{Pr}\left(\mathcal{D}_{1/1} \right)} + \frac{\mathrm{Pr}\left(\mathcal{D}_{1/1} \right) - \mathrm{Pr}\left(\mathcal{D}_{1/1,2} \right)}{\mathrm{Pr}\left(\mathcal{D}_{1/1} \right)\mathrm{Pr}\left(\mathcal{D}_{2/1,2} \right)}\lambda_2 < 1  , \lambda_2 < \mathrm{Pr} \left(\mathcal{D}_{2/1,2} \right)  \right\rbrace
\end{align}
\begin{align} \label{eq:R_2}
\mathcal{R}_2 = \left\lbrace (\lambda_{1},\lambda_{2}): \frac{\lambda_2}{\mathrm{Pr}\left(\mathcal{D}_{2/2} \right)} + \frac{\mathrm{Pr}\left(\mathcal{D}_{2/2} \right) - \mathrm{Pr}\left(\mathcal{D}_{2/1,2} \right)}{\mathrm{Pr}\left(\mathcal{D}_{2/2} \right)\mathrm{Pr}\left(\mathcal{D}_{1/1,2} \right)}\lambda_1 < 1  , \lambda_1 < \mathrm{Pr} \left(\mathcal{D}_{1/1,2} \right)  \right\rbrace
\end{align}
\end{figure*} 
 
\subsection{Second Dominant System: the second queue transmits dummy packets}

In the second dominant system, when the second queue empties then the source transmits a dummy packet for the $D_2$ while the first queue behaves in the same way as in the original system. In this dominant system, the second queue never empties, so the service rate for the first queue is 
\begin{eqnarray} \label{eq:mu1_D2}
\mu_1 = \mathrm{Pr}\left(\mathcal{D}_{1/1,2}\right).
\end{eqnarray}

The first queue is stable if and only if $\lambda_1 < \mu_1$. The probability that $Q_1$ is empty is
\begin{eqnarray} \label{eq:Pr1empty_D2}
\mathrm{Pr}\left(Q_1 = 0 \right)  = 1-\frac{\lambda_1}{\mathrm{Pr}\left(\mathcal{D}_{1/1,2}\right)}.
\end{eqnarray}

The service rate of the second queue, after substituting (\ref{eq:Pr1empty_D2}) into (\ref{eq:mu2}) is
\begin{equation} \label{eq:mu2_D2}
\mu_2 = \mathrm{Pr}\left(\mathcal{D}_{2/2}\right) - \frac{\mathrm{Pr}\left(\mathcal{D}_{2/2}\right) - \mathrm{Pr}\left(\mathcal{D}_{2/1,2}\right)}{\mathrm{Pr}\left(\mathcal{D}_{1/1,2}\right)}\lambda_1.
\end{equation}

The stability region $\mathcal{R}_2$ obtained from the second dominant system is given in (\ref{eq:R_2}).

\begin{figure}[]
\centering
\includegraphics[scale=0.6]{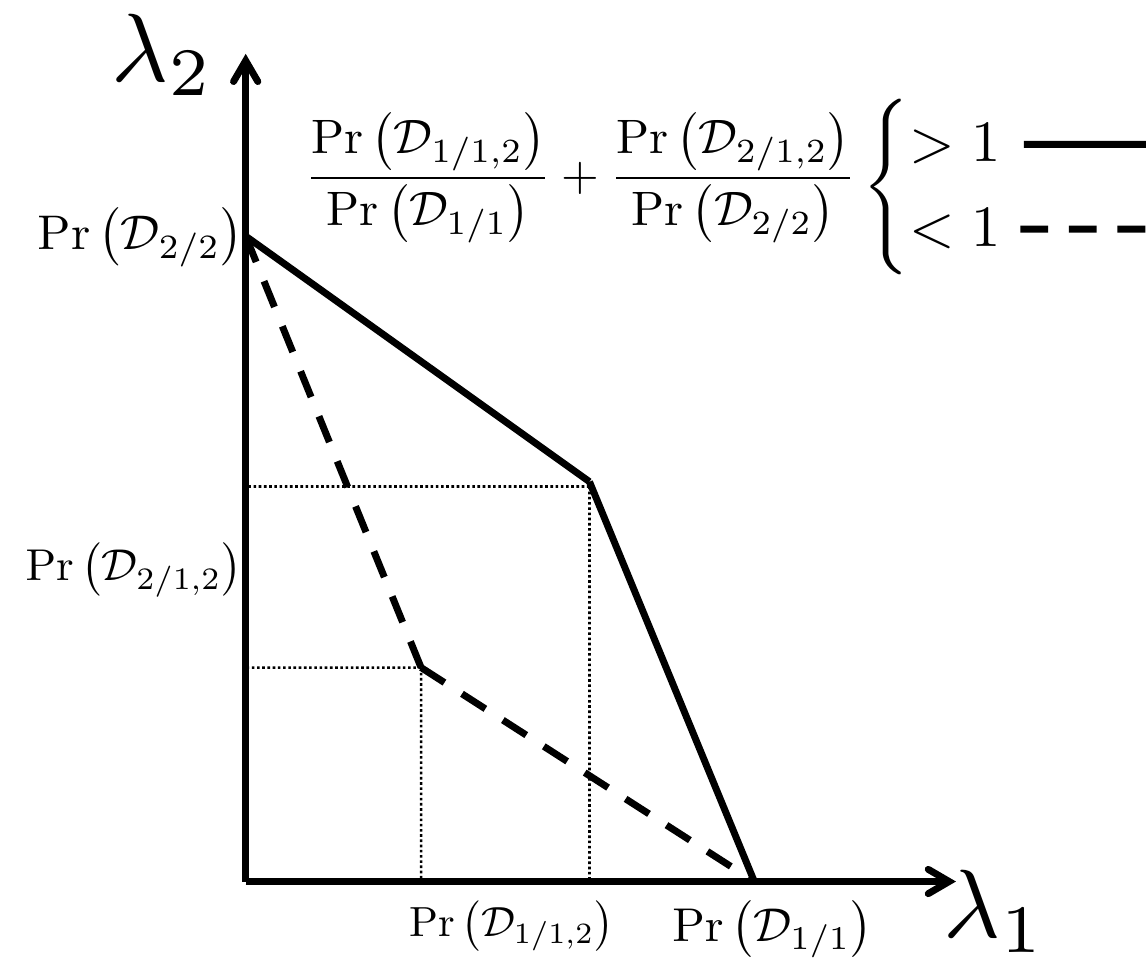}
\caption{The stability region for the two-user broadcast channel in the general case.}
\centering
\label{fig:region_general}
\end{figure}

The stability region of the system is given by $\mathcal{R} = \mathcal{R}_1 \bigcup \mathcal{R}_2$, where 
$\mathcal{R}_1$ and $\mathcal{R}_2$ are given by (\ref{eq:R_1}) and (\ref{eq:R_2}) respectively and is depicted by Fig. \ref{fig:region_general}.

An important observation made in \cite{rao:stability} is that the stability conditions obtained by the stochastic dominance technique are not only sufficient but also necessary conditions for the stability of the original system. The \emph{indistinguishability} argument~\cite{rao:stability} applies to our problem as well. Based on the construction of the dominant system, it is easy to see that the queues of the dominant system are always larger in size than those of the original system, provided they are both initialized to the same value. Therefore, given $\lambda_{2}<\mu_{2}$, if for some $\lambda_{1}$, the queue at $S_1$ is stable in the dominant system, then the corresponding queue in the original system must be stable. Conversely, if for some $\lambda_{1}$ in the dominant system, the queue at node $S_1$ saturates, then it will not transmit dummy packets, and as long as $S_1$ has a packet to transmit, the behavior of the dominant system is identical to that of the original system because dummy packet transmissions are eliminated as we approach the stability boundary. Therefore, the original and the dominant system are indistinguishable at the boundary points.

The obtained stability region for the two-user broadcast channel in the general case has the same expression with the stability region of the two-user interference channel obtained in \cite{b:pappasitw2013-ic}.

\section{Treating Interference as Noise}  \label{sec:stability_IAN}
In this section, we consider the case where the users treat the interfering signal as noise. When the $i$-th queue is empty at the source, while the $j$-th queue is not,  
then the success probability for the $i$-th user is given by (\ref{eq:SNR}). When both queues are non-empty then the transmitted signal at timeslot $t$ from the source to the receivers is denoted by $x^t=x_1^t+x_2^t$. The received signal $y_i^t$ by the user $D_i$ is 
$y_i^t = h_i^t x^t + n_i^t$. The event $\mathcal{D}_{i/i,j}$ denotes that user $D_i$ is able to decode its intended packet. This is feasible when the received SINR is above a threshold $\gamma_i$ and is expressed by

\begin{equation} \label{eq:SINR_IAN}
\mathcal{D}_{i/i,j}= \left\lbrace \frac{P_i \abs{h_i}^2 d^{-\alpha}_{i} }{1+P_j\abs{h_i}^2 d^{-\alpha}_{i}} \geq \gamma_i \right\rbrace.
\end{equation}

The success probability of the second user can be obtained similarly to the first. 
The transmission from the source to $D_2$ is successful when 
\begin{eqnarray*}
\frac{P_2 \abs{h_2}^2 d^{-\alpha}_{2} }{1+P_1\abs{h_2}^2 d^{-\alpha}_{2}} \geq \gamma_2  \iff \gamma_2 \leq \abs{h_2}^2 d^{-\alpha}_{2} (P_2 -\gamma_2 P_1) \\
\iff \gamma_2 \leq \abs{h_2}^2 d^{-\alpha}_{2} (P_2 -\gamma_2 (P-P_2)).
\end{eqnarray*}
Note that $P_1+P_2=P$. Thus, if $P_2 -\gamma_2 (P-P_2) < 0$ then, the success probability is zero because the initial inequality is not feasible. 
Thus, if $P_2 > \frac{\gamma_2}{1+\gamma_2}P$ then
\begin{eqnarray*}
\abs{h_2}^2 \geq \frac{\gamma_2}{d^{-\alpha}_{2} (P_2 -\gamma_2 (P-P_2))}.
\end{eqnarray*}
Assuming Rayleigh block fading, we have $\abs{h_2}^2 \sim \exp(1)$ and the success probability can be expressed as
\begin{equation}
\begin{aligned}
\mathrm{Pr} \left(\mathcal{D}_{2/1,2} \right)=\mathrm{Pr} \left[\abs{h_2}^2 \geq \frac{\gamma_2}{d^{-\alpha}_{2} (P_2 -\gamma_2 (P-P_2))}  \right] = \\ = \int_0^{\infty} \! \mathrm{Pr} \left[x \geq \frac{\gamma_2}{d^{-\alpha}_{2} (P_2 -\gamma_2 (P-P_2))}\right] f_{\abs{h_2}^2}(x)  \,  \mathrm{d}x.
\end{aligned}
\end{equation}

Thus,
\begin{equation}
\begin{aligned}
\mathrm{Pr} \left(\mathcal{D}_{2/1,2} \right)= \\ \int_0^{\infty} \! \left[ 1 - F_{\abs{h_2}^2} \left(x \geq \frac{\gamma_2}{d^{-\alpha}_{2} (P_2 -\gamma_2 (P-P_2))}\right) \right] f_{\abs{h_2}^2}(x)  \, \mathrm{d}x.
\end{aligned}
\end{equation}
Note that $f_{\abs{h_2}^2}(x)= \exp(-x)$ and $F_{\abs{h_2}^2} (x) = 1-\exp(-x)$.

To summarize, the success probability for the second user when both queues at the source are non-empty is given by (\ref{eq:Pr212}), where $\mathbbm{1} \{\cdot\}$ is the indicator function.

\begin{figure*}
\begin{align} \label{eq:Pr212}
\mathrm{Pr}\left(\mathcal{D}_{2/1,2} \right) = \mathbbm{1}\left\{ P_2 > \frac{\gamma_2}{1+\gamma_2}P \right\} \exp \left(- \frac{\gamma_2 d_2 ^{\alpha}}{ (P_2 -\gamma_2 (P-P_2)) } \right) = \mathbbm{1}\left\{ P_2 > \gamma_2 P_1 \right\} \exp \left(- \frac{\gamma_2 d_2 ^{\alpha}}{ P_2 -\gamma_2 P_1 } \right)
\end{align}
\end{figure*}

Thus if $P_2 > \gamma_2 P_1$ and $P_1 > \gamma_1 P_2$ then $\mathrm{Pr}\left(\mathcal{D}_{1/1,2} \right)=\exp \left(- \frac{\gamma_1 d_1 ^{\alpha}}{ P_1 -\gamma_1 P_2 } \right)$ and $\mathrm{Pr}\left(\mathcal{D}_{2/1,2} \right)=\exp \left(- \frac{\gamma_2 d_2 ^{\alpha}}{ P_2 -\gamma_2 P_1 } \right)$ after replacing in (\ref{eq:R_1}) and (\ref{eq:R_2}) we obtain the stability region $\mathcal{R} = \mathcal{R}_1 \bigcup \mathcal{R}_2$.
 
Similarly we can obtain the region for the other cases. 

\section{Successive Decoding}  \label{sec:stability_SC}
In this section we consider the case where the channel from the transmitter to $D_1$ is better than that to $D_2$;  i.e. $\abs{h_1} > \abs{h_2}$. When only one queue at the transmitter is non-empty, then the procedure of a successful transmission is described in Section \ref{sec:model}. When both queues at the source are non-empty, the procedure of decoding a packet by a receiver is as follows.

We refer to the two packets used in a single superposition-based transmission as two levels (layers). The packet intended for the weaker receiver (i.e. $D_2$)
is referred to as the first level. We refer to the other level as the second level. A transmitter using superposition coding splits the available transmission power between the two level, selects the transmission rate for each of the levels, then encodes and modulates each of the packets separately at the selected rate. The modulated symbols are scaled appropriately to match the chosen power split and summed to obtain the transmitted signal. More details about implementation of superposition coding at the medium access layer can be found in \cite{b:SCMobicom2007}.

At the receiver side, $D_2$ treats the message of $D_1$ as noise and decodes its data from $y_i$. Receiver $D_1$, which has a better channel, performs successive decoding, i.e. it decodes first the message of $D_2$, then it subtracts it from the received signal, and afterwards decodes its message with a single-user decoder. Note that in the broadcast channel with superposition coding, the decoding order is different from SIC in the interference channel, in which the signal with the strongest channel is decoded first. The successive decoding is feasible at the first receiver if 

\begin{eqnarray}
\left\lbrace \frac{P_2 \abs{h_1}^2 d^{-\alpha}_1} {1+P_1\abs{h_1}^2 d^{-\alpha}_{1}} \geq \gamma_2 ,\text{ } P_1  \abs{h_1}^2 d^{-\alpha}_{1} \geq \gamma_1 \right\rbrace.
\end{eqnarray}

$D_2$ is able to decode its intended packet if and only if the received signal-to-interference-plus-noise ratio (SINR) is greater than $\gamma_2$. 

The success probability seen by the first user, $D_1$ when both queues are non-empty is given by (\ref{eq:Pr112_SC}). The proof is omitted due to space limitations.

\begin{figure*}
\begin{align} \label{eq:Pr112_SC}
\mathrm{Pr}\left(\mathcal{D}_{1/1,2} \right) = \mathbbm{1}\left\{ \gamma_2 P_1 < P_2 \leq  P_1 \frac{\gamma_2 (1+\gamma_1)}{\gamma_1} \right\} \exp \left(- \frac{\gamma_2 d_1 ^{\alpha}}{ P_2 -\gamma_2 P_1 } \right) + \mathbbm{1}\left\{ P_2 > P_1 \frac{\gamma_2 (1+\gamma_1)}{\gamma_1} \right\} \exp \left(- \frac{\gamma_1 d^{\alpha}_{1}}{P_{1}}\right)
\end{align}
\end{figure*}

The probability that the link between the transmitter and $D_2$ is not in outage when both queues are non-empty is given by (\ref{eq:Pr212}) which is obtained in the previous section.

In the remainder, we consider two simple schemes regarding the transmission power for each receiver's packets. The first scheme is the case where we have fixed transmit power $P_i$ for the $i$-th receiver, such that $P_1+P_2=P$. The second scheme comes naturally whenever a user is inactive, i.e. has no packets to receive. We consider that the transmitter adapts the power considering the queue state of each receiver, i.e. if the queue $Q_i$ is empty, then all power $P$ is allocated to the $j$-th queue, ($i \neq j$).

\subsection{Fixed Power Scheme}
We assume here that the transmitter assigns fixed power $P_1$ (resp. $P_2$) at the $D_1$ (resp. $D_2$) on every timeslot.

\subsubsection{The case where $P_2 > P_1 \frac{\gamma_2 (1+\gamma_1)}{\gamma_1}$}
The service rate seen by the first queue is given by (\ref{eq:mu1}).
Since constant transmitting power $P_1$ is used and $D_1$ has better channel than $D_2$, from (\ref{eq:Pr112_SC}) we have that $\mathrm{Pr}\left(\mathcal{D}_{1/1,2} \right) = \mathrm{Pr} \left(\mathcal{D}_{1/1} \right)$. Thus, we have
\begin{eqnarray}
\mu_1 = \mathrm{Pr} \left(\mathcal{D}_{1/1} \right).
\end{eqnarray}

From Loyne's criterion for stability~\cite{b:Loynes}, the first queue is stable if and only if $\lambda_1 < \mu_1$.
From Little's theorem (Ch. 3.2 in~\cite{b:Bertsekas}), we have that
\begin{eqnarray} \label{eq:Prnonempty_fix}
\mathrm{Pr}\left(Q_1 > 0 \right) = \frac{\lambda_1}{\mathrm{Pr}\left(\mathcal{D}_{1/1} \right)}.
\end{eqnarray}

The service rate for the second queue is given by (\ref{eq:mu2}). After substituting (\ref{eq:Prnonempty_fix}) into (\ref{eq:mu2}) we obtain
\begin{eqnarray}
\mu_2 = \mathrm{Pr}\left(\mathcal{D}_{2/2} \right) + \frac{\mathrm{Pr}\left(\mathcal{D}_{2/1,2} \right) - \mathrm{Pr}\left(\mathcal{D}_{2/2} \right)}{\mathrm{Pr}\left(\mathcal{D}_{1/1} \right)}\lambda_1.
\end{eqnarray}

\begin{figure}[]
\centering
\includegraphics[scale=0.5]{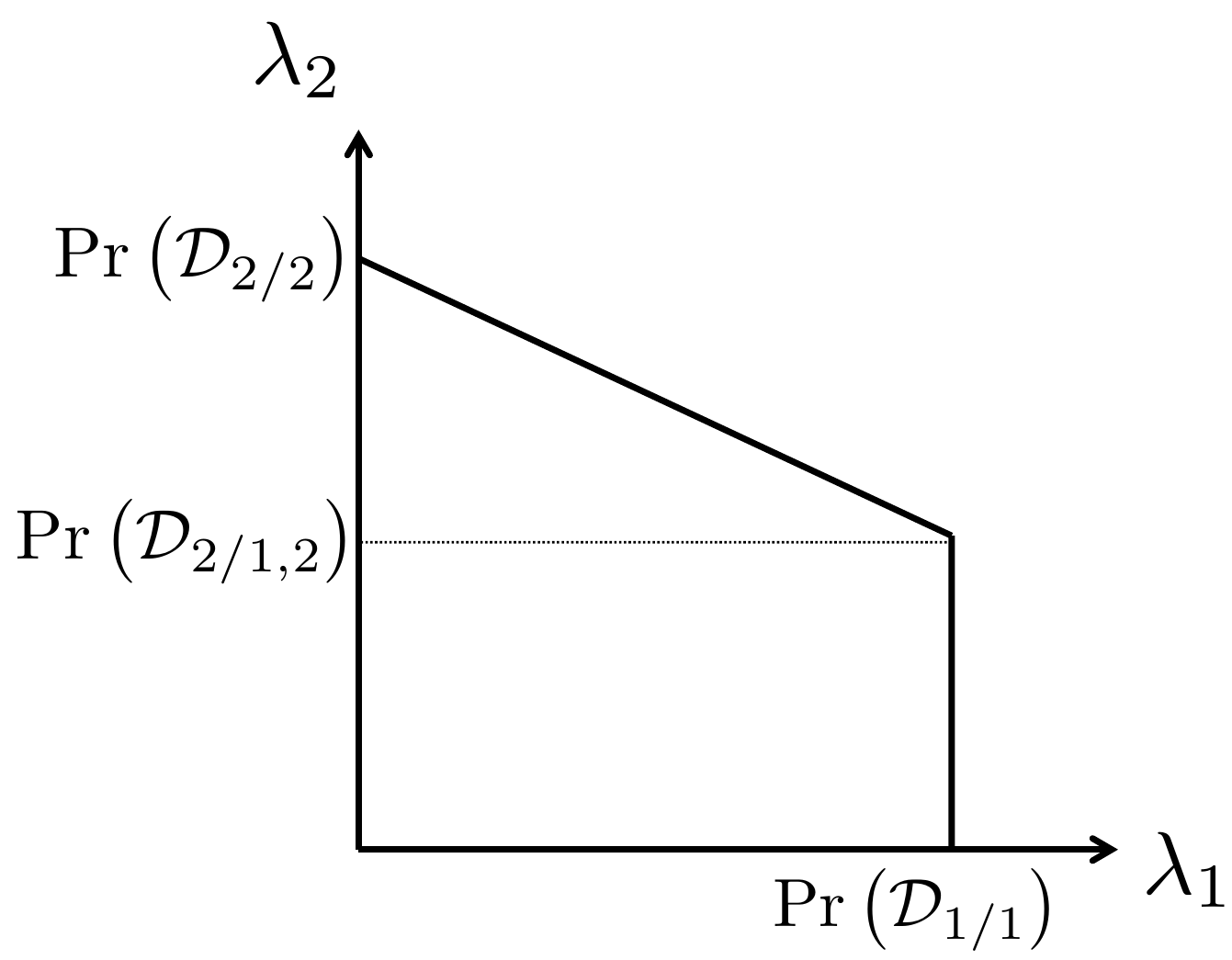}
\caption{The stability region for fixed transmit powers when $P_2 > P_1 \frac{\gamma_2 (1+\gamma_1)}{\gamma_1}$ and $D_1$ applies successive decoding and $D_2$ treats interference as noise. }
\centering
\label{fig:region}
\end{figure}

From Loyne's criterion we have that the second queue is stable if and only if $\lambda_2 < \mu_2$. The stability region for the degraded broadcast channel is given by (\ref{eq:Rfix}) is depicted by Fig.~\ref{fig:region}.

Recall that the success probability $\mathrm{Pr}\left(\mathcal{D}_{2/1,2} \right)$ is given by (\ref{eq:Pr212}).
Note that in this case we do not face the problem of coupled queues as mentioned in the general case described in Section \ref{sec:stability_general}.

\subsubsection{The case where $\gamma_2 P_1 < P_2 \leq  P_1 \frac{\gamma_2 (1+\gamma_1)}{\gamma_1}$}
In this case clearly $\mathrm{Pr}\left(\mathcal{D}_{1/1,2} \right) \neq \mathrm{Pr} \left(\mathcal{D}_{1/1} \right)$. We obtained that
\begin{equation} \label{eq:P112_case1}
\mathrm{Pr}\left(\mathcal{D}_{1/1,2} \right) = \exp \left(- \frac{\gamma_2 d_1 ^{\alpha}}{ P_2 -\gamma_2 P_1 } \right).
\end{equation}

In this case the queues are coupled so, we have to use the results obtained in Section \ref{sec:stability_general} derived by the stochastic dominance technique. After replacing (\ref{eq:P112_case1}) and (\ref{eq:Pr212}) into (\ref{eq:R_1}) and (\ref{eq:R_2}) we
obtain the stability region.

\begin{figure*}
\begin{align} \label{eq:Rfix}
\mathcal{R} = \left\lbrace (\lambda_{1},\lambda_{2}): \frac{\lambda_2}{\mathrm{Pr}\left(\mathcal{D}_{2/2} \right)} + \frac{\mathrm{Pr}\left(\mathcal{D}_{2/2} \right) - \mathrm{Pr}\left(\mathcal{D}_{2/1,2} \right)}{\mathrm{Pr}\left(\mathcal{D}_{1/1} \right)\mathrm{Pr}\left(\mathcal{D}_{2/2} \right)}\lambda_1 < 1  , \lambda_1 < \mathrm{Pr} \left(\mathcal{D}_{1/1} \right)  \right\rbrace
\end{align}
\begin{align}
\label{eq:R_1_changing}
\mathcal{R}_1= \left\lbrace  (\lambda_{1},\lambda_{2}): \frac{\lambda_1}{\exp \left( - \frac{\gamma_1 d_1^{\alpha}}{P} \right)} + \frac{\exp \left( - \frac{\gamma_1 d_1^{\alpha}}{P} \right) - \exp \left( - \frac{\gamma_1 d_1^{\alpha}}{P_1} \right)}{\exp \left( - \frac{\gamma_1 d_1^{\alpha}}{P} \right) \exp \left(- \frac{\gamma_2 d_2 ^{\alpha}}{ (1+\gamma_2) P_2 - \gamma_2 } \right)}\lambda_2 < 1, 
 \lambda_2 <  \exp \left(- \frac{\gamma_2 d_2 ^{\alpha}}{ (1+\gamma_2) P_2 - \gamma_2 } \right) \right\rbrace
\end{align}
\begin{align}
\label{eq:R_2_changing}
\mathcal{R}_2= \left\lbrace  (\lambda_{1},\lambda_{2}): \frac{\lambda_2}{\exp \left( - \frac{\gamma_2 d_2^{\alpha}}{P} \right)} + \frac{\exp \left( - \frac{\gamma_2 d_2^{\alpha}}{P} \right) - \exp \left( - \frac{\gamma_2 d_2^{\alpha}}{(1+\gamma_2) P_2 - \gamma_2} \right)}{\exp \left( - \frac{\gamma_2 d_2^{\alpha}}{P} \right) \exp \left(- \frac{\gamma_1 d_1 ^{\alpha}}{ P_1 } \right)}\lambda_2 < 1, \lambda_1 <  \exp \left(- \frac{\gamma_1 d_1 ^{\alpha}}{ P_1 } \right) \right\rbrace
\end{align}
\end{figure*}

\subsection{Variable Power Scheme based on Queue State}
In this part, we consider a simple adaptive scheme regarding the power allocation for each receiver's packets. 
The power allocation is performed as follows: when both queues are not empty, the transmit power for the first and second queue is $P_1$ and $P_2$, respectively, satisfying $P_1+P_2=P$. However, when the queue of $i$-th receiver is empty, the total transmit power $P$ is used for transmitting the packets from for the $j$-th (where $j \neq i$) receiver.

The average service rates of the first and the second queue, $\mu_1$, and $\mu_2$ are given by (\ref{eq:mu1}) and (\ref{eq:mu2}) respectively.
The success probabilities $\mathrm{Pr}\left(\mathcal{D}_{i/i} \right)$ for $i=1,2$ are given by
\begin{eqnarray} \label{eq:Prii}
\mathrm{Pr}\left(\mathcal{D}_{i/i} \right) = \exp \left( - \frac{\gamma_i d_i^{\alpha}}{P} \right),
\end{eqnarray}
since when a queue is empty, the transmitter assigns all power to the other queue, and can be obtained from (\ref{eq:SNR}). The success probability $\mathrm{Pr}\left(\mathcal{D}_{1/1,2} \right)$ is given by (\ref{eq:Pr112_SC}).

In the above scheme, it is evident that $\mathrm{Pr}\left(\mathcal{D}_{1/1} \right) \neq \mathrm{Pr}\left(\mathcal{D}_{1/1,2} \right)$, and as a result, there is coupling between the queues. Thus we can use directly the stability region obtained in Section \ref{sec:stability_general} by replacing the success probabilities.

The stability region $\mathcal{R}$ has two parts, $\mathcal{R}_1$ and $\mathcal{R}_2$ where $\mathcal{R} = \mathcal{R}_1 \bigcup \mathcal{R}_2$.
If $P_2 > P_1 \frac{\gamma_2 (1+\gamma_1)}{\gamma_1}$, then $\mathcal{R}_1$ is given in (\ref{eq:R_1_changing}) after replacing (\ref{eq:Pr212}), (\ref{eq:Pr112_SC}) and (\ref{eq:Prii}) into (\ref{eq:R_1}) , similarly $\mathcal{R}_2$ given by (\ref{eq:R_2_changing}). The stability region can be obtained similarly for the case $\gamma_2 P_1 < P_2 \leq  P_1 \frac{\gamma_2 (1+\gamma_1)}{\gamma_1}$.

The indistinguishability argument mentioned in Section \ref{sec:stability_general} applies to this case as well.

\section{Conclusions}
In this work, we derived the stability region for the two-user broadcast channel. We considered two decoding schemes at the receiver side, namely treating interference as noise by both receivers and successive decoding by the strong receiver. For the latter, two simple power allocation policies were studied, a fixed power allocation and an adaptive power scheme based on the queues' states.

\bibliographystyle{IEEEtran}
\bibliography{thesis}

\end{document}